\magnification\magstep1
\font\cst=cmr10 scaled \magstep3
\font\cstit=cmsltt10 scaled \magstep3

\vglue 1.5cm

\centerline{\cst On the mass of a Kerr-anti-de Sitter} 
\vskip 0.5cm
\centerline{\cst spacetime in {\cstit D} dimensions}

\vskip 1 true cm
\centerline{{\bf Nathalie Deruelle}$^*$ and {\bf
Joseph Katz$^{**}$}}
\vskip 0.5cm
\centerline{\it $^*$Institut d'Astrophysique de Paris,}
\centerline{\it GReCO, FRE 2435 du CNRS,}
\centerline{\it 98 bis boulevard  Arago, 75014, Paris, France}
\centerline{and}
\centerline{\it Institut des Hautes Etudes Scientifiques,}
\centerline{\it 35 Route de Chartres, 91440, Bures-sur-Yvette, France}
\bigskip
\centerline{\it$^{**}$ Institute of Astronomy, Madingley Road, Cambridge CB3 0HA,
UK}
\centerline{and}
\centerline{\it  The Racah Institute of Physics, Safra Campus, 91904
Jerusalem, Israel}\footnote{}{deruelle@ihes.fr

jkatz@phys.huji.ac.il}

\medskip
\vskip 0.8cm
\centerline{28 October 2004}
\vskip 0.5cm

\noindent {\bf Abstract}

\bigskip  We show how to compute the  mass  of a Kerr-anti-de Sitter spacetime
with respect to the anti-de Sitter background in any dimension, using a 
superpotential  which has been derived from standard N\oe ther identities. The
calculation takes
 no account of the source of the curvature  and confirms  results obtained  for black
holes via   the first law of thermodynamics.

\vskip .5 cm
\centerline{*}
\vskip .5 cm 

In an interesting paper Gibbons, Lu, Page and Pope [1] generalized the Kerr-(anti)-
de Sitter metrics to all dimensions $D>5$. Furthermore Gibbons, Perry and Pope [2]
calculated the angular momenta $J_{(D)i}$ of these rotating spacetimes using Komar's
integrals. They also calculated the mass $E_{(D)}$ of rotating {\it black holes} in anti-de
Sitter backgrounds using the first law of thermodynamics.  

Let us emphazise that
$E_{(D)}$ and $J_{(D)i}$ are classical concepts on which thermodynamics is built rather
than the other way round. A Kerr-anti-de Sitter spacetime with the same mass might
be produced by a rotating ``star" rather than a black hole. Such a body has no
quantum radiation nor a Bekenstein-Hawking entropy. One should surely be able to
calculate the mass in that case because the spacetime admits a timelike Killing
vector. 

Here  we show how to calculate the mass and the angular momenta  of 
Kerr-anti-de Sitter metrics   using the  covariant KBL superpotential [3] (see  [4]  
about the uniqueness and reliability of that superpotential).   

\vskip 0.7cm

 We choose to write the $D$-dimensional Kerr-anti-de Sitter metrics in Kerr-Schild
coordinates. In odd dimensions $D=2n+1$ for instance, the line element reads
$$
 ds^2=g_{\mu\nu}dx^\mu dx^\nu=d\bar s^2+{2m\over U}\,(h_\mu dx^\mu)^2\qquad
{\rm with}\quad \mu=\{0, 1, \mu_i, \phi_i\}\quad (i=1,...,n)
\eqno{(1)}
$$  where $d\bar s^2$ is the metric of the anti-de Sitter background and where the
coordinates $\mu_i$ are subject to the constraint
$\sum_{i=1}^{i=n}\mu_i^2=1$.  The mass parameter $m$ is an integration constant, 
$h_\mu$ is some null vector (denoted $k_\mu$ in [1-2]) depending on
$n$ rotation parameters $a_i$ and  $U$ is a scalar function. Explicit forms of those
functions can be found in Section 2 of ref [1]. These coordinates are the most
convenient for our calculations. The background admits a timelike Killing vector
$\xi$ and $n$ plane rotation Killing vectors $\eta_i$ associated with one parameter
displacements, say $\tau$.
$\xi$ is normalized in such a way that $\xi^\mu\delta\tau$   is an infinitesimal time
translation in a locally free falling {\it non rotating} inertial frame in the background.
Thus for the background metric in its conventional static form when
$D=2n+1$ for instance (equations (1.11) and (1.12) of Ref [1])
$$
 d\bar s^2= -\left(1+ {y^2\over l^2}\right)dt^2+{dy^2\over{1+{y^2/ l^2}}}+y^2 
d\sigma^2_{(D-2)}$$
$$\qquad\hbox{where}\quad  d\sigma^2_{(D-2)}=
\sum_{i=1}^{i=n}(d\tilde\mu_i^2+\tilde\mu_i^2 d\phi_i^2) ~~~{\rm
with}~~~\sum\limits_{i=1}^{i=n}\tilde\mu_i^2 =1,
\eqno{(2)}
$$ 
 the components of the timelike Killing vector corresponding to time translations
are  $ (1,0,0,0,...)$. In Kerr-Schild ellipsoidal coordinates
$\xi$ has   the same components. On the other hand, in  Boyer-Linquist coordinates
used in Section 3 of ref [1], the components are
$ (\xi^0=1,\xi^1=0,\xi^{\mu_i}=0,\xi^{\phi_i}=-a_i/l^2)$.

\vskip 0.5cm

We now turn to the calculation of $E_{(D)}$  using  the KBL superpotential 
$\hat J^{[\mu\nu]}$ (where a hat means multiplication by
$\sqrt{-g}$ and brackets denote antisymmetrization) which is defined as (see [5] for a
recent review and references)~:
$$
\hat  J^{[\mu\nu]}= -{1\over 8\pi}\left(D^{[\mu}\hat\xi^{\nu]}-
\overline{D^{[\mu}\hat\xi^{\nu]}}+\hat\xi^{[\mu}k^{\nu]}\right)
\eqno{(3)}
$$
with a ``Newton constant" $G_D\equiv 1$. The first term
in (3) is the well known Komar superpotential density of the foreground with metric
$\bf g$. The second term substracts the Komar superpotential of the background
with metric
$\bf{\bar g}$. The vector $k^\nu$ in the third term is~:
$$ k^\nu =
g^{\nu\rho}(\Gamma^\sigma_{\rho\sigma}-\overline{\Gamma^\sigma_{\rho\sigma}})-g^{\rho\sigma}(\Gamma^\nu_{\rho\sigma}-\overline{\Gamma^\nu_{\rho\sigma}})\,.
\eqno{(4)}
$$ The mass $E_{(D)}$ is given by the integral of $\hat  J^{[\mu\nu]}$  on a
$(D-2)$ surface $S$ of radius $x^1=r\rightarrow \infty$ in a $(D-1)$ hypersurface
$x^0=t=const.$ in Kerr-Schild coordinates :
$$
 E_{(D)}= \int_S d^{D-2}x\,\hat J^{[01]}\,.\eqno{(5)}
$$  We repeat that the integrand being covariant,  the integral is a scalar which may
be  calculated in any convenient coordinates.  

Angular momenta $J_{(D)i}$ are obtained by
replacing $\xi$ by
$\eta_i$ in the KBL superpotential $\hat J^{[01]}$. Since
$\eta_i^0=\eta_i^1=0$ it reduces to the Komar superpotentials and :
$$ J_{(D)i}=  \int_S d^{D-2}x\hat J_i^{[01]}\quad\hbox{with}\quad \hat
J_i^{[01]}=-{1\over 8\pi}
\left(
D^{[0}\hat\eta_i^{1]}-\overline{D^{[0}\hat\eta_i^{1]}}\right)\,.
\eqno{(6)}
$$ 

\vskip 0.8cm

The calculation of $\hat J^{[01]}$ for the mass is a mechanical procedure, which is
straightforward in Kerr-Schild coordinates\footnote{$^*$}{The useful properties
are that the metric coefficients do not depend on $x^0$ and
$\phi_i$; the $g_{\mu_i\mu_j}$ coefficients do not depend on $m$; the other
off-diagonal components are linear in $m$ and hence do not enter the calculation.
The fact that the vector $h^\mu$ is null further simplifies the calculation since
$\sqrt{-g}=\sqrt{-\bar g}$.}. Using the asymptotic form of the metric given in [1] at
leading order in $r$ we arrive at :
$$
 \hat J^{[01]}={m\over 8\pi\Xi}\,[(D-1)W-1]\sqrt{g_{(D-2)}}+{\cal
O}\left({m^2\over r}\right)\,,
\eqno(7)
$$ 
where $g_{(D-2)}$ is the determinant of  the metric $d\sigma^2_{(D-2)}$ in (2) with tildes
dropped and
 $$
\Xi\equiv\prod\limits^{i=n}_{i=1}\Xi_i
~~~,~~~W\equiv\sum\limits_{i=1}^{i=n}{\mu_i^2\over\Xi_i}~~~,~~~
\Xi_i\equiv 1-{a_i^2\over l^2}~,
\eqno(8)
$$
with $a_n=0$ in even dimensions $D=2n$.

As for the calculation of $\hat J_i^{[01]}$ for the angular momenta  it yields (the
background term $\overline{D^{[0}\hat\eta_i^{1]}}$ does not contribute)
$$
\hat J_i^{[01]}= {m\over
8\pi\Xi}{a_i\mu_i^2\over\Xi_i}(D-1)\sqrt{g_{(D-2)}}+{\cal O}\left({m^2\over
r}\right)\,.
\eqno(9)
$$ 

While the ``multi-cylindrical"  pairs of coordinates $(\mu_i, \phi_i)$ were
convenient in [1]  to find the Kerr-anti-de Sitter metric as well as in our previous
calculations, the explicit integration of (5)(7) and (6)(9) is more
easily performed
 in spherical coordinates, say, $\varphi_k$ with $(k=1, 2,\cdots,
D-2)$. The result is
$$
E_{2n}=m{{\cal V}_{2n-2}\over
4\pi\Xi}\sum\limits_{i=1}^{i=n-1}{1\over\Xi_i}~~~,~~~
E_{2n+1}=m{{\cal V}_{2n-1}\over
4\pi\Xi}\big(\sum\limits_{i=1}^{i=n }{1\over\Xi_i}-{1\over 2}\big )~~~,~~~
J_{(D)i}=m{{\cal V}_{D-2}\over 4\pi\Xi}{a_i\over\Xi_i}
\eqno(10)
$$
where ${\cal V}_N$ is the volume of a $N$-sphere.

These results confirm those given in [2]. However they were obtained  using
standard ideas about global N\oe ther conservation laws associated with
translational and rotational invariance of a  background spacetime.They depend on
the asymptotic form of the metric only and hence take no account of the source of
the curvature. In [2] the mass was obtained by  using the first law of black hole
thermodynamics given that the entropy is identified with
$1/4$ of the black hole area and the    temperature with  
 its surface gravity.  Here we know the mass in the first place and assuming the same
temperature as in [2] we may  compute  the entropy of a black hole if there is one.
 
\vskip .5 cm
\noindent {\bf Acknowledgments}
\medskip N.D. thanks the Cambridge Centre for Mathematical Sciences and the
Yukawa Institute for Theoretical Physics in Kyoto for their  hospitality and
acknowledges financial support from a CNRS-Royal Society Exchange Program.
J.K. is grateful for the usual hospitality and financial support of the Institute of
Astronomy in Cambridge where most of his work was performed.
 
\vskip .5 cm
\noindent {\bf Note Added in Proof}
\medskip
Since this work was completed, Gary Gibbons, Malcolm Perry and Chris Pope
showed that the mass can also be computed by means of the
Ashtekar-Magnon-Das conformal boundary formula ; see version (2) of
their paper [2].

\vskip .5cm

\noindent {\bf References}
\medskip
\item{[1]} G.W. Gibbons, H. Lu, Don N. Page and C.N. Pope, {\it The General Kerr-de
Sitter Metrics in All Dimensions}, hep-th/0404008.
\item{[2]} G.W. Gibbons, M.J. Perry and C.N. Pope, {\it The First Law of
Thermodynamics for Kerr-de Sitter Metrics}, hep-th/0408217.
\item{[3]} J. Katz, {\it A note on Komar's anomalous factor},  Class. Quantum Grav.,
{\bf 2} (1985) 423; J. Katz, J. Bi\v c\'ak, D. Lynden-Bell, {\it Relativistic conservation
laws and integral constraints for large cosmological perturbations}, Phys. Rev. {\bf
D55} (1997) 5759.
\item{[4]} B. Julia and S. Silva, {\it Currents and superpotentials in classical gauge
theories~: II. Global aspects and the extension of affine gravity}, Class. Quantum
Grav., {\bf 17} (2000) 4733, gr-qc/0005127.
\item{[5]} N. Deruelle, J. Katz and S. Ogushi, {\it Conserved Charges in Einstein-
Gauss-Bonnet theory}, Class. Quantum Grav., {\bf 21} (2004) 1971, gr-qc/0310098.
\end